\documentclass[twocolumn]{grant_aastex}

\usepackage{verbatim}
\usepackage{epstopdf}
\usepackage{apjfonts}
\usepackage{amsmath}
\usepackage{our_macros}

\shortauthors{O'Dea et al.}
\shorttitle{X-ray gas around CSS Radio Galaxies}

\begin{document}

\title{Testing for shock-heated X-ray gas around Compact Steep Spectrum Radio Galaxies}

\correspondingauthor{Christopher P.~O'Dea}
\email{chris.odea@umanitoba.ca}

\author{C.~P.~O'Dea}
\affiliation{Department of Physics \& Astronomy, University of Manitoba, Winnipeg, MB R3T 2N2, Canada}
\affiliation{School of Physics \& Astronomy, Rochester Institute of Technology,
84 Lomb Memorial Drive, Rochester, NY 14623, USA}

\author{D.~M.~Worrall}
\affiliation{HH Wills Physics Laboratory, University of Bristol, Tyndall Avenue,  Bristol\ \ BS8 1TL, UK}

\author{G.~R.~Tremblay}
\affiliation{Einstein Fellow, Department of Astronomy and Yale
Center for Astronomy and
Astrophysics, Yale University, 52 Hillhouse Ave.,
New Haven, CT 06511, USA}
\affiliation{Harvard-Smithsonian Center for Astrophysics,
60 Garden St., Cambrdige, MA 02138, USA}

\author{T.~E.~Clarke}
\affiliation{Naval Research Laboratory Remote Sensing Division, Code 7213
4555 Overlook Ave SW, Washington, DC 20375, USA}

\author{B.~Rothberg}
\affiliation{Large Binocular Telescope Observatory, 933 North Cherry Ave., Tucson, AZ 85721, USA}

\author{S.~A.~Baum}
\affiliation{Carlson Center for Imaging Science,  Rochester Institute of Technology,
84 Lomb Memorial Drive, Rochester, NY 14623, USA}
\affiliation{Faculty of Science, University of Manitoba, Winnipeg, MB R3T 2N2, Canada}

\author{K.~P.~Christiansen}
\affiliation{InsightSTEM, 855 E River Rd \#37, Tucson, AZ 85718 USA}

\author{C.~A.~Mullarkey}
\affiliation{School of Physics \& Astronomy, Rochester Institute of Technology,
84 Lomb Memorial Drive, Rochester, NY 14623, USA}

\author{J.~Noel-Storr}
\affiliation{InsightSTEM, 855 E River Rd \#37, Tucson, AZ 85718 USA}
\affiliation{Designated Campus Colleague, Steward Observatory, University of Arizona, 933 N Cherry Ave,
Tucson AZ 85719}

\author{R.~Mittal}
\affiliation{Center for Computational Relativity and Gravitation, Rochester Institute of Technology,
170 Lomb Memorial Drive, Rochester, NY 14623, USA}

\begin{abstract}
	We present  \chandra\ and \xmm\ X-ray, VLA radio, and optical observations of
	three candidate Compact Steep Spectrum (CSS) radio galaxies. CSS  sources are galactic scale
	and are presumably driving a  shock through the ISM of their host galaxy.
	\source\ is a low excitation emission line CSS radio galaxy with
	possibly a hybrid Fanaroff-Riley FRI/II (or Fat Double) radio morphology. The \chandra\
	observations  reveal a point-like source which is  well fit with a power law consistent with
	emission from a Doppler boosted  core. 
	  3C\,268.3 is a CSS broad line
	 radio galaxy whose \chandra\ data are consistent spatially
    with a point source centered on the nucleus and spectrally with a double power-law model. 
	\xsource\ is a low excitation emission line radio galaxy with a bent
	double radio morphology.   While from our new spectroscopic redshift, PKS B1017-325 falls outside the
       formal definition of a CSS, the  \xmm\ observations are consistent  with ISM emission with either a contribution from hot shocked gas or non-thermal jet emission.
	 We compile selected radio and X-ray
	properties of the nine {\it bona fide\/} CSS radio galaxies with X-ray detections so far. We find
	that 2/9 show X-ray spectroscopic evidence for hot shocked gas. We note that the counts in the
	sources are low and the properties of the 2 sources with evidence for hot
	shocked gas are typical of the other CSS radio galaxies. We suggest that hot
	shocked gas may be typical of CSS radio galaxies due to their propagation
	through their host galaxies.
\end{abstract}


%%%%%%%%%%%%%%%%%%%%%%%%%%%%%%%%%%%%%%%%%%%%%%%%%%%%%%%%%%%%%%%%

\section{Introduction} \label{sec:intro}

The required gaseous medium for the propagation of extragalactic radio sources
is provided by the X-ray-emitting atmospheres of their host galaxies and clusters.
In radio sources of high power, the energy and momentum fluxes are expected to
be sufficient to drive a bow shock at supersonic speed into the ambient medium,
heating the gas as it crosses the shock \cite[e.g.,][]{scheuer, leahy, heinz}.
Evidence for such heating has been claimed around Cygnus\,A
\citep[e.g.,][]{carilli-cyga, smith-cygacluster}.  In cluster-embedded sources
of lower power it is common for the radio lobes to displace X-ray-emitting gas
\cite[e.g.,][]{boer-n1275rosat, birzan, forman, wise, tremblay12b}, with sufficient power to
balance radiative cooling of the dense gas \cite[e.g.][]{dunn,
rafferty,fabian2012}.  However, direct evidence for strong shock heating by
radio sources remains relatively scarce. Notable exceptions are the closest
radio galaxy, Centaurus~A  \citep{kraft-cenalobe, croston},  the cluster radio
source 3C444 \citep{croston2011}, 
and the
intermediate-power radio galaxy PKS\,B2152-699 \citep{wfos} where prominent
shocks have been studied in detail. There has been less routine success in
detecting the anticipated dramatic effects of shocks around high-power sources in more typical atmospheres.

A promising source type for observing shock heating is the non-cluster compact
steep spectrum (CSS) population, where the radio structures lie within the
volume of the host galaxy \citep[for a review see][]{odea}.  Here, as for the
inner expanding structures of Centaurus\,A, the ambient gas is the cool galaxy
atmosphere, which even when shocked should be accessible to detection in the
X-ray band.  While CSS sources can show indirect evidence of interaction with
the gas in their host galaxies \citep[e.g.,][]{saikia,Orienti2016}, the amount
of gas directly involved appears to be relatively small
\citep[e.g.,][]{axon,stockton}, and early suggestions that CSS sources are
compact because they are
``frustrated'' by the presence of dense environments are now generally
disfavored as compared with arguments that they are compact because they are
young \citep[e.g.,][]{fanti1995}.  Typical age estimates of less than $10^6$ years
are based on measuring curvature in the radio spectra caused by radiative energy
losses of the higher-energy electrons over the lifetime of the sources
\cite[e.g.,][]{murgia} or through proper motions of the terminal hotspots of the
class of smaller and younger Compact Symmetric Objects (CSOs, a subclass of GPS sources)
\cite[e.g.,][]{oc,
taylor, peck2000,polatidis,An2012}.  CSS radio sources are powerful, with
statistical arguments made that they decline in luminosity as they age
\citep[e.g.,][]{fanti1995, readhead}.  As such they are good candidates for
searching for shock heating of surrounding gas at moderate to high Mach number.

The compact nature of CSS sources, and their smaller cousins, Gigahertz Peaked
Spectrum (GPS) sources, has the disadvantage that the emission might
be confused with other components of X-ray emission.  Quasars, whose
X-ray cores and inner jets are enhanced by relativistic beaming, and
where the potential to image shock discontinuities is diminished
through the jets lying far from the plane of the sky, are
not ideal for such study.  Moreover, it has been argued from
polarization and multifrequency variability studies that quasar
samples of genuinely young sources can be contaminated by blazars
\citep{cotton, tornikoski}. Where radio
galaxies have their cores detected in X-ray and separated from other
X-ray components,  the nuclear emission is much fainter than
in quasars and correlated with the core radio emission (e.g., \citealt{wb94, evans}, and see
\citealt{worrall} for a review).  In CSS radio galaxies
the radio core emission is often too weak to be separated from the
steep-spectrum lobes, particularly at low radio frequencies where it is
self-absorbed, and core X-ray emission may
be weak compared with that from the gaseous environment or
possibly that associated with the steep-spectrum radio lobes.   CSS radio galaxies
 are therefore good candidates for observing shock heating.

%\begin{figure} \centering
%	\includegraphics[width=0.8\columnwidth,clip=true]{3c303_redo.pdf}
%	\caption{The short 8~ksec \chandra\ observation of 3C\,303.1 \citep{massaro2010}
%		contains 16 counts ($0.3-7$ keV), shown here in native $0\farcs492 \times 0\farcs492$ pixels.
%The distribution is more extended than the point-spread function, supporting conclusions of \citet{odea06}, based on deeper \xmm\ spectroscopy, that the emission is predominantly from galaxy-scale gas (2\arcsec corresponds to 8.2~kpc at the source's redshift of $z=0.267$).
%The grey-scale shows  $0.3-7$ keV counts. The radio emission is double with no obvious core, oriented NW-SE (similar to the X-ray emission, \citet{massaro2009}) with a largest angular size of 1.2 arcsec at 8.4~GHz \citep{akujor}.
%	} \label{fig:3c303.1} \end{figure}

While CSS and GPS quasars and broad-line radio galaxies are detected
quite readily in X-rays \citep[e.g.,][]{odea, worr48, guainazzi04,
siemiginowska05, siemiginowska08}, the same is not true for radio
galaxies.  After unsuccessful attempts to detect them with \rosat\
\citep{odea96}, the first detection of a compact radio galaxy was of
the GPS source B1345+125 ($z=0.122$) with \asca\ \citep{odea00}.  In
addition to B1345+125, we are aware of 32 other GPS radio galaxies
having reported detections with \xmm\ or \chandra\ \citep{vink,
	guainazzi06, siemiginowska08, tengstrand,Siemig2016}.   These GPS  sources all have
radio sizes less than 1~kpc (most are less than 100 pc) and, even with \chandra, structures on
the scale size of the radio cannot be resolved.  Count rates are typically low,
and all have been fitted to the simplest model of an absorbed
single-component power law.  \citet{tengstrand} demonstrates a
possible anticorrelation between radio-source size and X-ray intrinsic
absorption in GPS galaxies  (but see \citet{vink} for a different
conclusion), as would be consistent with the presence of an
inner absorbing torus as seen in powerful radio galaxies with
large-scale structures \citep[e.g.,][]{evans} and X-ray emission that
arises further from the nucleus in larger GPS sources.  This
interpretation is also consistent with the finding that GPS quasars
show no intrinsic absorption \citep{siemiginowska08}.  However, while
the location of absorption may be probed, the emission mechanism
responsible for the X-rays on various spatial scales is far from
certain.   \citet{Siemig2016} find that their results do not favor
thermal emission from the hot ISM for the X-ray continuum emission of the CSO
subset of GPS galaxies.

% and they may be associated with the presence of an accretion
%disk, arise from inverse Compton scattering in the compact radio structures,
%or be the emission from thermal gas.

Until relatively recently, there has been less effort to detect CSS radio galaxies in X-rays,
despite the fact that their radio structures are on spatial scales
resolvable with \chandra.  \citet{Kunert2014} observed
7 CSS sources with low radio power and detected four of them.
Three of the low radio power CSS sources have extended X-ray emission -
0810+077, 0942+355, and 1321+045 which is in a cluster of galaxies
\citep{Kunert2013}, though all of the
four detected sources were fit by an absorbed power-law.

The low redshift CSS radio galaxy 3C305 \citep{heckman1982} shows shocked hot
gas in the X-rays \citep{massaro2009,hardcastle2012}.  The hot gas seems likely
to be  participating in an outflow driven by the radio source which includes the
cool atomic and warm optical emission line gas
\citep{morganti2005,hardcastle2012,reynaldi2013}.  In such an outflow scenario,
the X-ray-emitting phase dominates the energetics of the (known) phases of the
gas in the outflow \citep{hardcastle2012}.

The high throughput of  \xmm\  is useful for the detection of CSS radio
galaxies, and we made a good detection of the $z=0.267$ galaxy 3C\,303.1 in a
40~ksec observation \citep{odea06}.  In the spectrum we found a thermal component
at $kT \sim 0.8$~keV, which seems likely to be produced by the interstellar
medium (ISM) of the host galaxy, and a second harder component.  The lack of a
detected radio core suggested to us that it is unlikely that the hard component
arises from the nucleus.  For the harder component to originate from
inverse-Compton scattering in the brighter radio lobe, the magnetic field would
have to be a factor of about 4 lower than the minimum-energy value.  We argued
that a plausible hypothesis for the second component is that it is gas that is
heated by the bow shock of the expanding radio source.  Using an expansion speed
of $\sim 6000$~km s$^{-1}$ as suggested by lifetime and cooling-time arguments,
particularly in the emission line nebula \citep{devries, odea02}, the shock would
be very strong (Mach number of 13) and the hot gas should have $kT \sim 43$~keV,
consistent with the observed hard spectral component.  3C\,303.1 has
subsequently been observed with \chandra\ for $8$ ksec as part of a shallow survey
of radio sources.  Examination of the archival data \citep{massaro2010} shows a
detection with only 16 photons, but consistent with expectations from our \xmm\
detection.  While this is insufficient to probe spectral structure, we find that
photons are distributed in a more extended fashion than would be the case for a
point source  \citep{massaro2010},    %(\autoref{fig:3c303.1}),
supporting our hypothesis that the
emission arises from gas rather than being non-thermal in nature.
%The emission of the GPS source B1345+125 is also
% reported as resolved with \chandra,  consistent with a hot-gas origin \citep{siemiginowska08}.

In this paper we report new X-ray observations of the
CSS radio galaxies  \source\ ($z=0.195$) and   3C\,268.3 ($z=0.3717$)
and candidate CSS source \xsource\
which is at a redshift of $z=0.318$ (based on new Large Binocular Telescope optical spectroscopy, see \autoref{sec:opticaloobs}). We adopt
$H_0 = 70$~km~s$^{-1}$~Mpc$^{-1}$, $\Omega_{m_0} = 0.3$,
$\Omega_{\Lambda_0} = 0.7$.  1~arcsec corresponds to 3.2, 5.1, and 4.6~kpc
at the distances of \source, 3C\,268.3, and \xsource, respectively.

\begin{deluxetable*}{llcccrccc}
	\tablecaption{Properties of X-ray detected CSS Radio Galaxies  \label{tbl-xray}}
	\tablewidth{0pt}
	\tablehead{
		\colhead{Sources} & \colhead{z }  & \colhead{Size (kpc)}  & \colhead{log P$_{\mathrm 1.4 GHz}$} & \colhead{log L$_{\mathrm X-ray}$}
		& \colhead{Counts}  & \colhead{Extended?} & \colhead{Shock?}  & \colhead{Refs}
	}
	\startdata
	B0810+077     & 0.1122 &   3.0 & 25.16 & 42.8$^a$ &    119 & Y & N & 1 \\
	B0942+355    & 0.2076  &    5.1  & 25.25 & 42.9$^a$  &    103  & Y & N & 1 \\
	3C237            & 0.877    &  9.2   & 28.3   & 44.0$^b$  & ...  & Y    & N  &  9 \\
	PKS B1017-325 & 0.3186 & 44$^a$  & 26.77 & 42.8$^c$  &  163   & N$^a$ & ? & 3 \\
	3C 268.3        & 0.3717 &  6.9 &  27.07 & 44.5$^b$ &    398 & N & N & 3,7 \\
	B1321+045    & 0.263  & 12 &  25.43 & 42.3$^a$ &    53 & Y & N & 1,2 \\
	3C303.1     &  0.2704 &   7.0 & 26.63 & 42.9$^b$ &  211 & Y & Y & 4,5 \\
	B3 1445+410 & 0.195 & 26 &  25.65 & 42.9$^c$ &   723  & N & N & 3 \\
	3C 305         & 0.0416  &  11 &  25.10 & 41.7$^b$ &    148 & Y & Y & 6,8 \\
	B1558+536   & 0.179 &  6.0 &  25.19 & 41.7$^a$  &    9 & N & N & 1 \\
	\enddata
	\tablecomments{Selected properties of CSS sources detected in X-rays. Col 1. Source name.   Col 2. redshift.
		Col 3. Largest linear size (e.g., diameter) of the radio source in kpc. $^a$Based on our new spectroscopic redshift, the candidate CSS source \xsource\ is now known to have a projected linear diameter of 44 kpc which is significantly above the cutoff for CSS sources. Thus, \xsource\ is not considered a CSS for
		the  analysis in \S \ref{sec:compare}.
		Col 4. Log Radio Power at rest frame 1.4 GHz (W Hz\mone). A radio spectral index of $\alpha = -0.75$ was adopted.  Col 5. X-ray Luminosity (erg s\mone). $^a$ 2-10 keV, $^b$ 0.5-7 keV,
		$^c$ 0.4 - 6 keV.
		Col 6. \chandra\ X-ray counts, except for 3C303.1 and PKS B1017-325 which are XMM counts. We use the  same energy band as used in Col 5.
		Col 7. Is there evidence for extended X-ray emission? $^a$The constraint on \xsource\ is based on XMM data which does not have sufficient angular resolution to rule out a galaxy-scale ISM.
		Col 8. Is there evidence for shocked hot gas in the X-rays?
		Col 9. References for columns 5-8: 1. \citet{Kunert2014}. 2. \citet{Kunert2013}. 3. This paper. 4. \citet{odea06}.
		5. \citet{massaro2010}. 6.  \citet{massaro2009}. 7. \citet{massaro2015}. 8. \citet{hardcastle2012}. 9.
\citet{massaro2017}.	}
\end{deluxetable*}

\section{Observations \& analysis}
\label{sec:obs}

\subsection{The Candidate CSS Sources }

The CSS \source\ is part of the B3-VLA sample of CSS sources defined by
\citet{fanti2001}. The CSS sources are a subset of the B3-VLA sample
\citep{vigotti1989} and have flux densities at 408 MHz $S_{408} > 0.8$ Jy, and
based on VLA A configuration 1.5 GHz observations have projected linear diameter
$\lae 28 $ kpc
and steep spectral index $\alpha \lae -0.5$ where $S_{\nu} \propto \nu^\alpha$.
The measured angular size of $\theta \sim 8\arcsec$
(\autoref{fig:1445taperradio}) gives a projected linear diameter of 26 kpc.

  3C\,268.3, z=0.3717, is classified optically as a BLRG. It is a member of the original \citet{fanti1990}
 sample of CSS sources. The projected total size of the radio source is 6.9 kpc. 

The  candidate CSS \xsource\ is a member of the Molonglo Southern 4 Jy Sample (MS4)
\citep{burgessa} and was one of 41 sources in MS4 classified as a CSS source by
\citet{burgessb} based on linear diameter $<25$ kpc (using their photometric redshift)
 and spectral index $\alpha <
-0.5$ between 2700 and 408 MHz. Our VLA radio image (Figure~\ref{fig:10172massradio}) shows an angular
size of  $\theta \sim 9.6\sec$ which corresponds to a projected linear diameter of 44
kpc using the new spectroscopic redshift and scaling. Thus, \xsource\ is actually larger than the cutoff for CSS sources.  We present the data for \xsource\ because it helps to provide constraints on the sizes of radio sources which are capable of  shocking the ISM. However, we do not include \xsource\ in the analysis of CSS sources in \S~\ref{sec:compare}.

\subsection{\chandra}
\label{sec:chandra}

We observed \source\ at the nominal aimpoint of the S3 CCD of the
\chandra\ ACIS camera on 2009 Sep 27.  The observations were in VFAINT
full-frame data mode (OBSID 11579).  Results presented here use CIAO
v4.8 and CALDB v4.7. We reprocessed the data following the software
`threads' from the \chandra\ X-ray Center
(CXC)\footnote{http://cxc.harvard.edu/ciao} to make a new level 2
events file.  The observation was free from background flares, and
after removal of time intervals when the background deviated more than
$3\sigma$ from the average value, the total exposure time is 68.541
ksec.  It was necessary to shift the X-ray coordinates by about 0.26
arcsec, mainly in RA, to align the nuclear emission with the radio
core at RA= 14:47:12.758, Dec=+40:47:45.00.  The 0.4-5 keV count rate
from a 1.25-arcsec radius circle at the core is 0.01 counts s$^{-1}$,
rendering pile-up negligible (1\% level).  The CIAO {\sc wavdetect}
task was run to find unassociated point sources that were then
excluded from the annular region used to measure the background for
our spectral fitting.
Spectral fitting was performed using {\sc xspec} version 12.9.0u
taking into account the absorption along the line of sight
in our Galaxy by using a column density of $N_{\mathrm H} = 1.3 \times
10^{20}$ cm$^{-2}$ (from the {\sc colden} program provided by the \chandra\
X-ray Center (CXC), using data of \citealt{dlock90}).  A
circle of radius 1.25 arcsec was used to extract a binned spectrum of
the core.

3C\,268.3 was observed with \chandra\ ACIS-S in VFAINT full-frame data mode
on 2009 July 29th and 30th (ObsIDs 10382 and 10933).  Initial results were
published by \citet{massaro2015}. We  reprocessed the data using CIAO 4.9 and CALDB 4.7.5.1, giving a total
observing time after background screening of 71.520 ks.  Galactic
absorption of $N_{\rm H} = 1.91\times 10^{20}$ cm$^{-2}$ was applied to all
spectral models.

\subsection{\xmm}
\label{sec:xmmobs}

We observed \xsource\ with \xmm\ on 2008 Jun 11   (ObsID 0552070101).
In this paper we
report on the data from the European Photon Imaging Camera (EPIC).
The pn, MOS1, and MOS2 CCD-array cameras were operated in Full Frame
mode with the thin optical blocking filter.  The tasks {\sc epchain}
and {\sc emchain} from the {\sc sas} were run to create events lists
(see http://xmm.esac.esa.int/ for information on the modes of
operation of the cameras and analysis procedures). There were no times
of extreme background flaring, and the resulting good exposure times,
after removing intervals when the background deviated by more than
3$\sigma$ from the average value, were 15.749, 15.800, and 12.832~ksec
in the MOS1, MOS2, and pn, respectively.  Our analysis uses good
events with patterns 0 to 12 from the MOS data and patterns 0 to 4
from the pn data.

For spectral analysis, background-subtracted events were extracted for
each camera separately using the task {\sc evselect}, and
corresponding calibration files were made using the {\sc rmfgen} and
{\sc arfgen} tasks from version 15 of the {\sc sas}. Counts were
extracted from a source-centered circle, with local background
measured from the same CCD.  For the pn we followed the recommended
procedure of taking background from the same distance to the readout
node as the source region.  For the MOS cameras this restriction was
dropped and somewhat larger background regions were used.  The
spectral extractions were fitted jointly to models using {\sc xspec}.
Data were grouped to 10 counts/bin and the weighting scheme of
\citet{churazov} was adopted to provide an improved estimate of the
variance in the limit of small numbers of counts.  Spectral fits
include absorption along the line of sight in our Galaxy assuming a
column density of $N_{\mathrm H} = 5.55 \times 10^{20}$ cm$^{-2}$ (from
the {\sc colden} program provided by the CXC using data of
\citealt{dlock90}).

\subsection{Radio}
\label{sec:radioobs}

\subsubsection{\source\ }

\begin{figure}
	\centering
	%emulateapj%
	\includegraphics[width=1.0\columnwidth,clip=true]{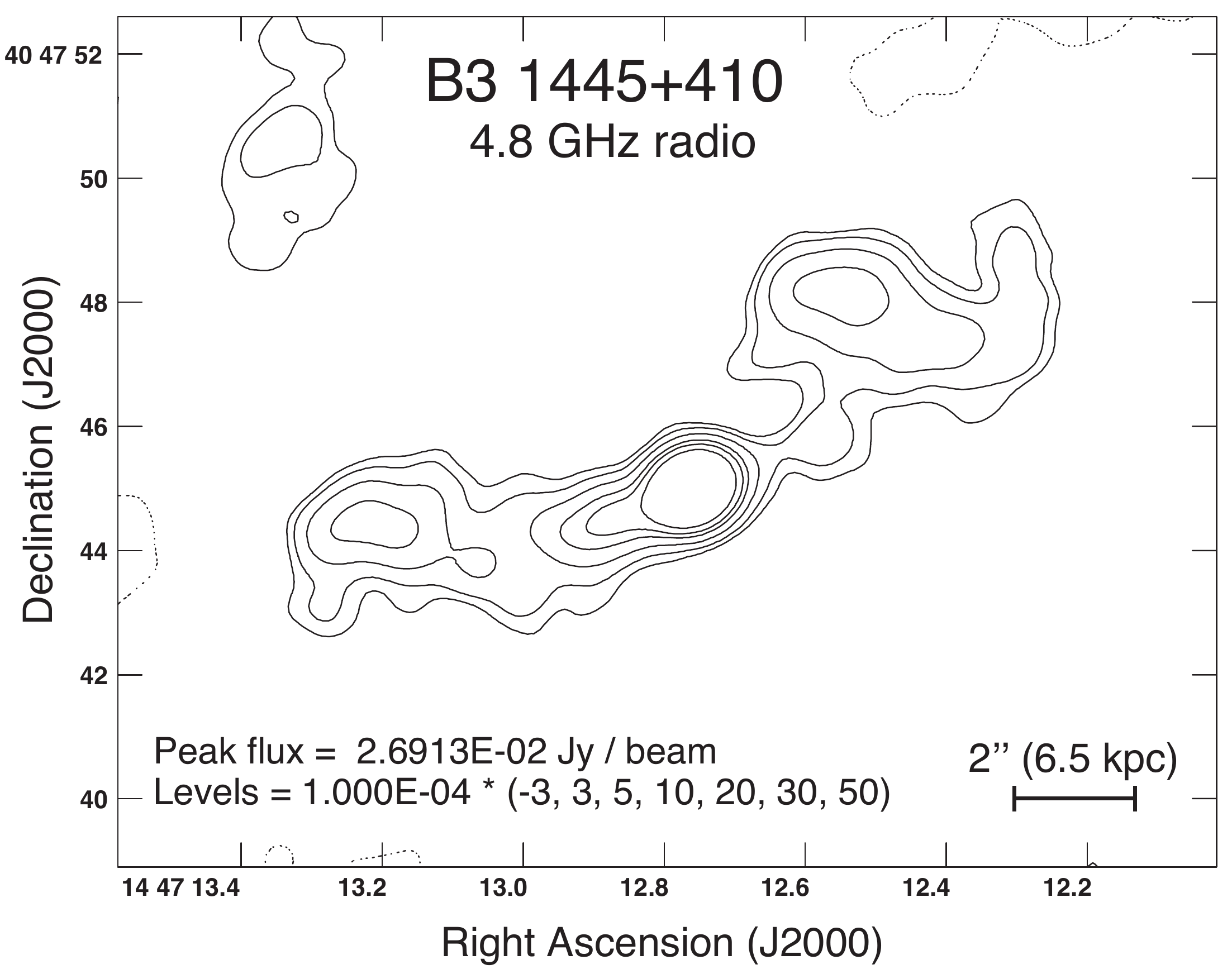}
	\caption{\source\ VLA 4.8 GHz image with UV data tapered to give a resolution of
		$0.91 \times 0.76 $ arcsec
		at PA $-62.0\deg$ in order to show intermediate scale structure.
		The radio contours are at -0.3, 0.3, 0.5,1.0,2.0,3.0,5.0 mJy beam$^{-1}$.
	}
	\label{fig:1445taperradio}
\end{figure}

We obtained archival 4.86 and 8.46 GHz A configuration
Karl G. Jansky Very Large Array (VLA) observations  
(Program AF0398, 2003-Jun-01) of \source. The observations were short snapshots - time on source 170 s at 8.4 GHz, and 130s at 4.8 GHz, respectively.  
The NRAO AIPS package was used for the calibration, imaging,
self-calibration, and deconvolution. The 8.46 GHz image (not shown) resolves out
nearly all of the extended source structure, however the radio nucleus is
detected with a flux density of $S_{8.5} \simeq 25$ mJy. The core is detected
in the 4.86 GHz image with a similar flux density of $S_{4.9} \simeq 26$ mJy,
giving a  flat spectral index for the core of $\alpha \simeq -0.03 $

\citet{fanti2001}  present 1.5 GHz VLA A configuration observations of \source\
which shows a double morphology. Our tapered 4.86 GHz image
(\autoref{fig:1445taperradio}) shows the extended structure in more detail.   Note that because of the limited uv coverage,  structures larger than $\sim 10\sec$ are not adequately imaged. 
The lobes have weak hot spots and there is a one-sided FRI-like jet extending
from the core into the eastern  lobe. This structure  suggests an intermediate
FRI/FRII morphology consistent with  the class of ``Fat Doubles'' described by
\citet{owen1989}.  The very weak hot spots and plume-like lobes suggest the
lobes are not expanding rapidly into the ISM/ICM.

\subsubsection{\xsource\ }

We observed \xsource\ with the NRAO VLA on 2011 Feb 01 using the hybrid CnB
array with extended north arm due to the southern declination of the target
(Program 10C-233). The observations covered a total block of 1.5 hours with
antenna pointing corrections made at 8.44 GHz and the target observed with a
total bandwidth of 256 MHz centered on 22.4 GHz. The phase calibrator used was
J1018$-$3123 and the flux/bandpass calibrator was 3C~286.

The data were calibrated and imaged with the NRAO Obit package \citep{Obit}
following standard procedures. Briefly, the data were Hanning smoothed to
suppress ringing due to radio frequency interference (RFI). They were flagged
for RFI that was impulsive in time or frequency. The calibrator 3C~286 was used
to determine corrections for instrumental group delay offsets which were applied
to all data. 3C~286 was also used to determine the bandpass to correct for
residual variations of gain and/or phase with frequency across the band. The
flux scale was set using 3C~286 and was bootstrapped to observations of the phase calibrator
J1018$-$3123, which bracketed the target observations. The phase calibrator was
used to calibrate the amplitudes and phases of our target. Following this
initial full calibration and flagging, we reset the calibration and undertook a
second full round of RFI flagging and calibration on only the data that passed
the initial calibration round. Calibration was applied to \xsource\ and it was
split off for imaging. The data were imaged and underwent 3 rounds of phase-only
self-calibration. The final image has an rms of 77.6 $\mu$Jy/beam with a beam of
1.1$\times0.6\arcsec\ $at a position angle of 48.4 degrees.

\begin{figure}
	\centering
	\includegraphics[angle=90,width=0.5\textwidth]{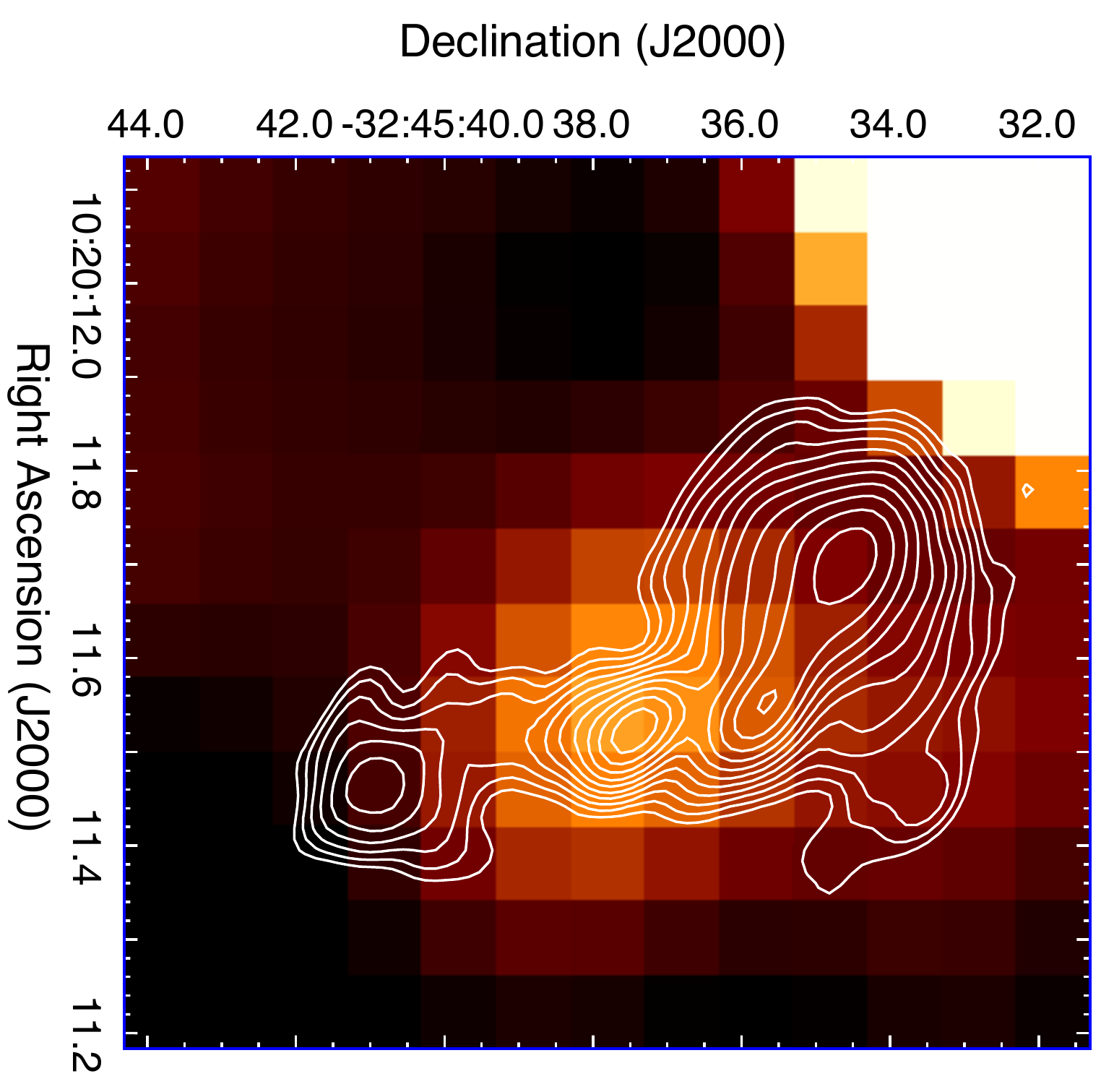}
	\caption{2MASS \citep{2MASS} K band image of \xsource\ with 22.4 GHz radio
          contours overlaid. The radio core is co-spatial with the
          host galaxy while a nearby star is seen to the
          north-east. Contours start at 5$\sigma$ and increase by
          $\sqrt{2}$, where $\sigma=77.6 \mu$Jy/bm.}
	\label{fig:10172massradio}
\end{figure}

The 22.4 GHz image (\autoref{fig:10172massradio}) shows an amorphous morphology
with a central `core' component that is co-spatial with an optical and IR point source
detected in ground-based imaging.  A single Gaussian component fit to the apparent core 
gives a position of  RA 10 20 11.52411   $\pm   0.0002879$
 DEC -32 45 37.4292   $\pm    0.004541$, where the uncertainty in RA and DEC is given in arcsec.
To the north of the
core there is a bright elongated region roughly  4.9 arcsec (22.5 kpc) in length
elongated to the north-east. In addition, there is a faint suggestion of further
emission extending $\sim$ 7 kpc to the west of the northern portion of the
bright emission. To the south of the core, there is faint emission connecting to
a brighter compact region. This emission extends 21.6 kpc to the south of the core. The
overall radio source morphology is unclear but might be  a ``bent double'' or
wide-angle-tail radio source  \cite[e.g.,][]{owen1976,blanton2001}. The total
flux density of all components of \xsource\ at 22.4 GHz is 110.0 $\pm$ 0.6 mJy with 14.2
$\pm$ 0.1 mJy in the core component. We note that Gaussian component fitting to
the `core' region reveals an extended structure so we are likely measuring a
combination of emission from the central AGN and surrounding region.

\begin{deluxetable}{c|cc}
	%\tabletypesize{\scriptsize}
	\tablecaption{Emission line properties of \xsource\ }
	\tablewidth{0pt}
	\tablehead{
		\colhead{} & \colhead{FWHM} & \colhead{Flux} \\
		\colhead{Line} & \colhead{(km s\mone)} & \colhead{($\times 10^{-15}$ erg s\mone\ cm\mtwo) }
	}
	\startdata
	$[$\ion{O}{2}$]$ $\lambda3726$ \AA &  627 $\pm$ 94   &  1.54 $\pm$ 0.23\\
	$[$\ion{Ne}{3}$]$ $\lambda3868$ \AA &   573 $\pm$ 85   &  0.214 $\pm$ 0.03 \\
	H$\gamma$ $\lambda$4340	\AA	 &   539 $\pm$ 80   &  0.576 $\pm$ 0.09\\
	\ion{He}{2} $\lambda$4685	\AA	 &  551 $\pm$ 83   &  0.427 $\pm$ 0.06\\
	H$\beta$ $\lambda$4861 \AA  &  656 $\pm$ 98  & 1.91 $\pm$ 0.18 \\
	$[$\ion{O}{3}$]$ $\lambda4959$ \AA &  660 $\pm$ 99 & 4.87  $\pm$ 0.73\\
	$[$\ion{O}{3}$]$ $\lambda5007$ \AA &  665 $\pm$ 98 & 14.8 $\pm$ 2.2 \\
	$[$\ion{N}{1}$]$ $\lambda5198$ \AA &  337 $\pm$ 51    & 0.162  $\pm$ 0.02 \\
	$[$\ion{O}{1}$]$ $\lambda6300$ \AA &   491 $\pm$ 74   & 0.852 $\pm$ 0.13  \\
	$[$\ion{O}{1}$]$ $\lambda6363$ \AA &   302 $\pm$ 45   & 0.285 $\pm$ 0.04 \\
	$[$\ion{N}{2}$]$ $\lambda6548$ \AA &   658 $\pm$ 99   &  1.99 $\pm$ 0.30 \\
	H$\alpha$ $\lambda6563$ \AA  &  602 $\pm$ 90 & 8.11 $\pm$ 1.2\\
	$[$\ion{N}{2}$]$ $\lambda6583$ \AA &   736 $\pm$ 111   &  6.50 $\pm$ 0.98\\
	$[$\ion{S}{2}$]$ $\lambda6716$ \AA &  418 $\pm$ 63    &  2.78 $\pm$ 0.42\\
	$[$\ion{S}{2}$]$ $\lambda6731$ \AA &  418 $\pm$ 63    &  2.06 $\pm$ 0.31\\
	\enddata
	\tablecomments{Emission line properties from the new LBT optical spectrum acquired for \xsource. Only those lines
	detected at $\gae5\sigma$ are reported. The redshift of all detected lines reported above is $z=0.318 \pm 0.009$. The excitation
	index as defined by \citet{buttiglione2010} is EI$\sim$0.63, consistent with the source being a Low Excitation Galaxy (LEG). }
	\label{tab:pks1017_spectrum}
\end{deluxetable}

\subsection{Optical Spectroscopy}
\label{sec:opticaloobs}

\subsubsection{\source}

\begin{figure*}
	\centering
	\includegraphics[width=1.0\textwidth]{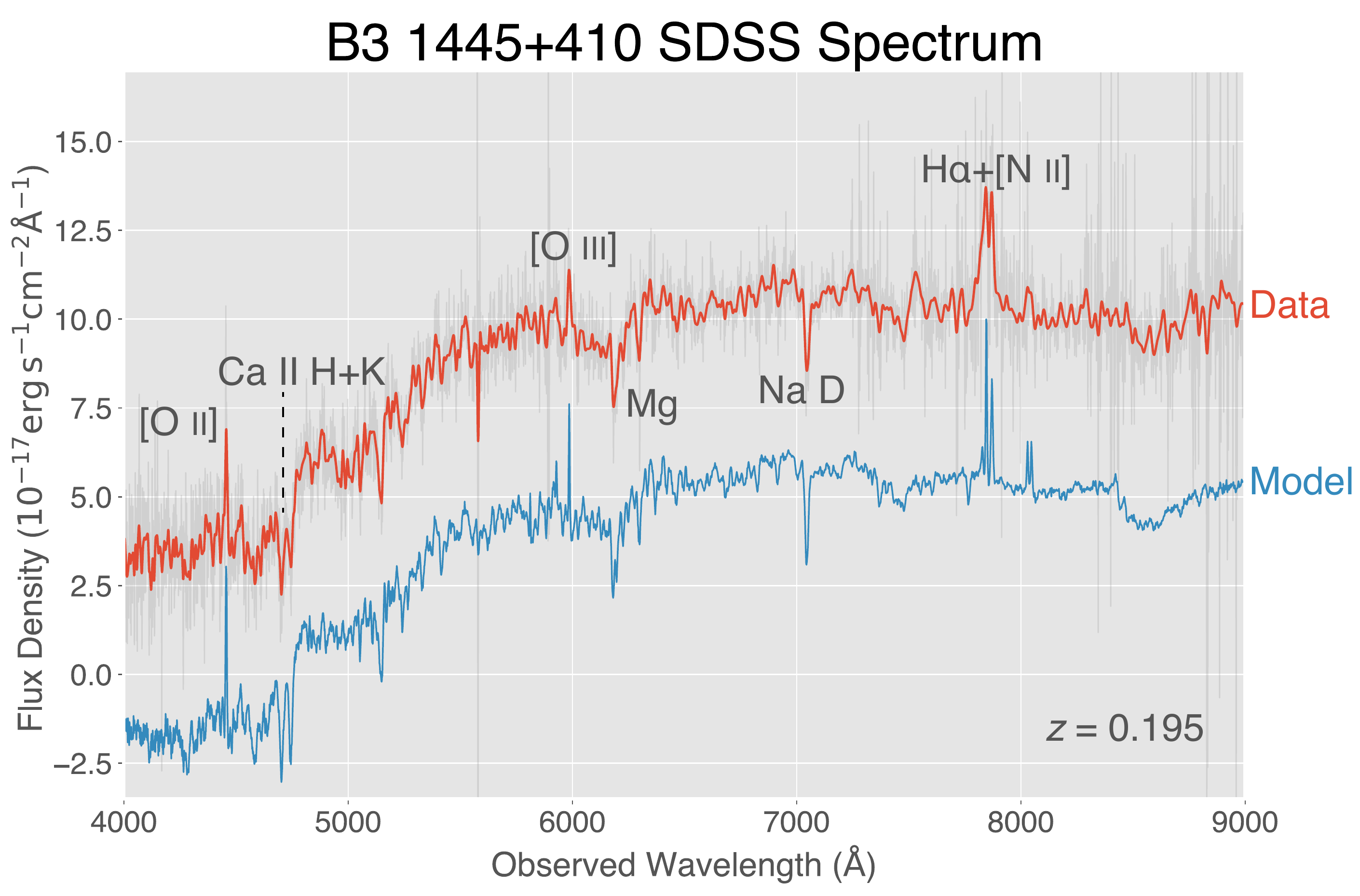}
	\caption{\source\ SDSS spectrum showing a red elliptical galaxy with low ionization emission lines.
	}
	\label{fig:1445optical}
\end{figure*}

The Sloan Digital Sky Survey (SDSS)  spectrum of \source\ is shown in
\autoref{fig:1445optical}. The SDSS observations were taken %in ???, using
from plate 1397 and fiber 182. The datasets were used as delivered  from the SDSS
archive, with no post-processing applied. The spectrum is that of a red
elliptical galaxy with low ionization emission lines, e.g., H$\alpha$+[NII],
[OII]$\lambda 3727$, and H$\beta$. Mg and Na are seen in absorption. The
redshift is $z = 0.1952 \pm 0.0002$.
%The properties of the emission lines are given in Table XX.
The excitation index EI\footnote{\citet{buttiglione2010} defines the excitation index
 EI$=$ log $[$\ion{O}{3}$]$/H$\beta$ - 1/3 (log $[$\ion{N}{2}$]$/H$\alpha$ +
  log $[$\ion{S}{2}$]$/H$\alpha$ + log $[$\ion{O}{1}$]$/H$\alpha$)}
\citep{buttiglione2010}
is $EI \sim 0.6$,
consistent with \source\ being a Low Excitation Galaxy
(LEG, also as defined by \citealt{buttiglione2010}).

\subsubsection{Large Binocular Telescope Optical Spectroscopy of \xsource}

\begin{figure*}
	\centering
	\includegraphics[width=1.0\textwidth]{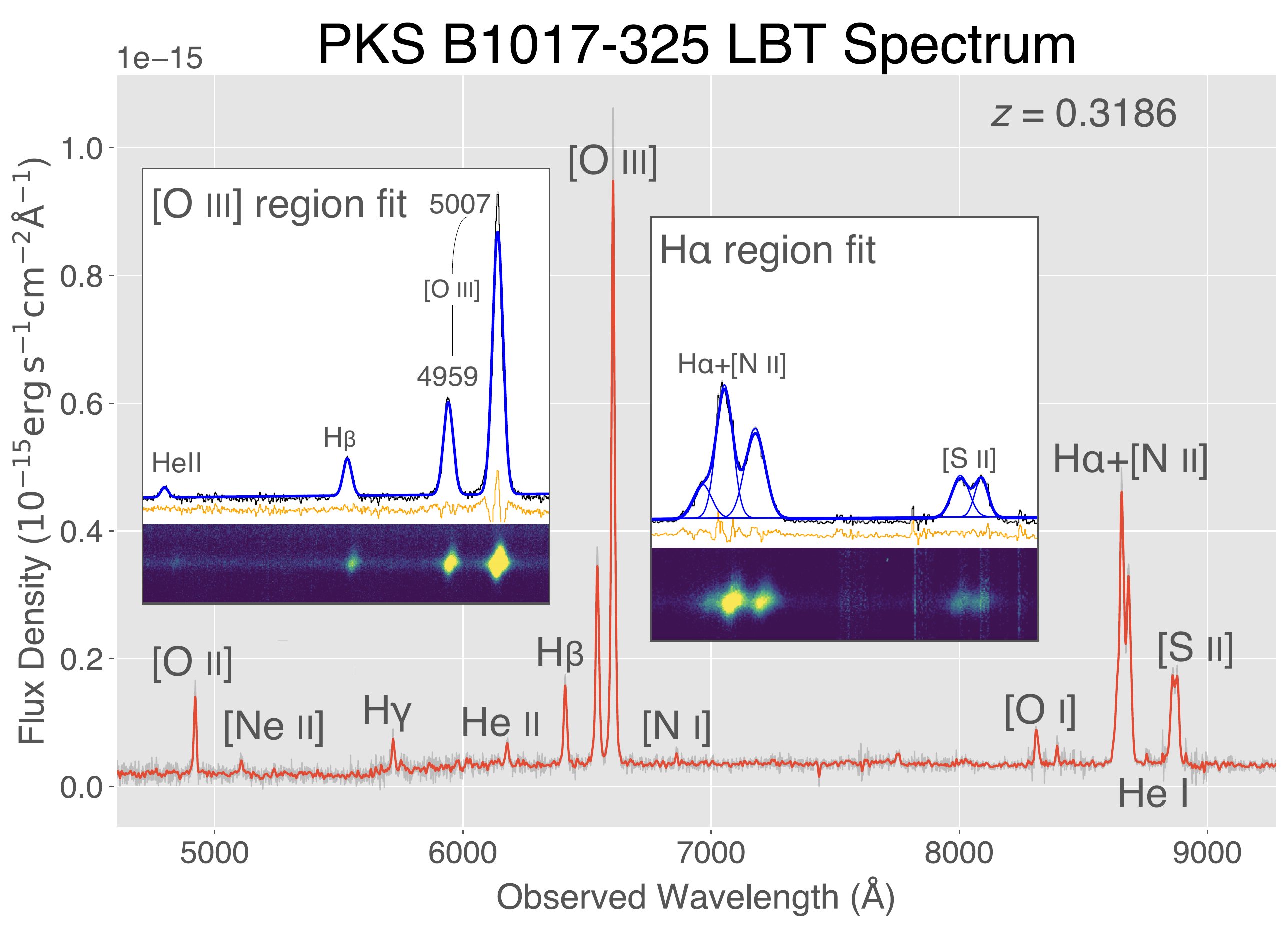}
	\caption{The new LBT/MODS2 optical spectrum of \xsource.
	The inset panels show the results of simultaneous multi-component
	Gaussian fits to emission lines in the $[$\ion{O}{3}$]$ and $H\alpha$
	regions of the spectrum. In those inset panels, the data and errorbars
	are shown in black and gray (respectively), our composite
	multi-component fit to the data is shown in blue, and the residual
	between that fit and the data is shown in orange. All detected
	lines in the spectrum are consistent with a redshift of $z=0.3186\pm0.009$.
	This, the first spectroscopic redshift
	yet obtained for \xsource, is inconsistent with the photometric
	redshift reported for the source in NED.
	}
	\label{fig:1017optical}
\end{figure*}

An optical spectrum of \xsource\ (\autoref{fig:1017optical})
was obtained on UT January 11, 2017
using the  Multi-Object Double Spectrograph 2~\citep[MODS-2;][]{pogge2006}.
MODS-2 is one of  two identical optical low-medium resolution two channel (blue \&
red) spectrographs  mounted on the Large Binocular Telescope (LBT). MODS-1 and
MODS-2 are each mounted  on one of the two f/15 straight Gregorian foci of the
two 11.8 meter mirrors.  MODS-2 is located on the DX (or right) side of the
telescope.  Although the LBT has moved to nearly full-time binocular duplex
(identical configuration)  observations with all three of its facility
instruments, including MODS 1 \& 2   \citep{rothberg}, PKS 1017-325
could only be observed in monocular mode primarily due to the low
declination of the target.

In standard  Gregorian or non-prime focus
operations, LBT uses an adaptive secondary  (AdSec) mirror with each primary
mirror  \citep{gallieni,quiros,miller}.
Each AdSec employs a thin shell 91.1cm mirror with 672 actuators. These mirrors
are limited in operation to elevations above 30$^\circ$ even when used in
seeing limited (non-adaptive) mode. LBT has a single rigid secondary mirror that
is not subject to the same limits. As PKS 1017-325 never reaches an elevation
above 30$^\circ$ at Mount Graham, only the rigid secondary, which was
already mounted on the DX side of the telescope,  could be used.\\

\indent
Either MODS spectrograph can be figured to simultaneously observe the wavelength
range  0.3-1.0\micron\ using a dichroic to split the light into a blue and red
channel.  MODS-2 was  configured in the dual grating mode using a
$1{\arcsec}\hspace{-1mm}.2$ wide segmented longslit  mask.  The mask contains 5
longslit segments, each 60{\arcsec} in length.  A single segment was  sufficient
for the observations.  The spatial scale of the blue arm and red arms are
$0{\arcsec}\hspace{-1mm}.120$~pixel$^{-1}$ and
$0{\arcsec}\hspace{-1mm}.123$~pixel$^{-1}$, respectively.  The observations were
obtained under nearly Full Moon (98.7$\%$) conditions with  consistent thick
clouds and seeing of $\sim$ 2{\arcsec} as determined from the acquisition
image. A single blue and red exposure of 1200 seconds was taken of PKS 1017-325.
The spectrophotometric standard G191B2b was observed using the same instrumental
setup as the  science observations, with the exception of the slitwidth. A
5{\arcsec} wide slit was used in order to minimize light loss for flux
calibrations. Slitless pixel flats were obtained to  correct for the detector
response. Ar, Xe, Kr, Ne and Hg lamps were observed for wavelength
calibration.\\  \indent The first step of the data reduction was to use Version
2.0 of the modsCCDRed collection  of Python scripts. These perform basic 2D
reduction including: bias subtraction, removing the  overscan regions,
constructing flat fields from the slitless pixel flats, fixing bad columns and
flipping the red arm data so that wavelength increases from left to right along
the x-axis.   Further processing was performed using customized \textsc{IRAF}
scripts developed by B.~Rothberg  \citep{secrest}.  Briefly,
cosmic rays were removed using the task \texttt{CRUTIL}.   Two dimensional
spectra were extracted in strip mode for the central longslit segment using the
task  \texttt{APALL}.  MODS spectra are tilted along both the spectral (x-axis)
and spatial (y-axis) dimensions.   The spectra were simultaneously corrected in
both axes using the arc lines and the \textsc{IRAF}  tasks \texttt{ID},
\texttt{REID}, \texttt{FITC}, and \texttt{TRANSFORM}.  MODS data are very
sensitive  to the polynomial order used to wavelength calibrate the data. A 4th
order Legendre polynomial produces  the smallest residuals and avoids
introducing low-order noise into the 2D spectra. Once the 2D spectra  were
rectified and wavelength calibrated, the background was subtracted in each
exposure by fitting a  2nd order Legendre polynomial to the columns
(spatial axis) using the \textsc{iraf} task  \texttt{BACKGROUND}.  A 1D spectra
was extracted in a metric aperture of 2{\arcsec} for both the blue and red
exposures.  The  exposures were then flux calibrated using the
spectro-photometric  standard G1919B2b. This step also removes the instrumental
signature.  The red data (0.56-1.0$\mu$m)  were corrected for telluric
absorption features using a normalized spectra of G191B2b in conjunction  with
the \textsc{IRAF} task \texttt{TELLURIC}.  The data were then corrected for
atmospheric extinction  and Galactic reddening assuming $R_\textrm{V}=3.1$ and a
value of $A_\textrm{V}=0.178$~\citep{schlafly2011}. Finally, the 1D blue and red
spectra were combined using the \textsc{IRAF} task \texttt{SCOMBINE}, corrected
to a heliocentric velocity, rescaled to a common dispersion value of 0.85~{\AA}
pixel$^{-1}$,  and trimmed to a wavelength range of 0.38-0.9285 $\mu$m.
Measurements of the sky lines yields a spectral  resolution of 4.2{\AA} in the
blue and 7.3{\AA} in the red, which corresponds to a resolution of \ {\it R}
$\sim$ 1300 across the entire spectral range.

Simultaneous Gaussian fits to the $[$\ion{O}{3}$]~\lambda\lambda$4959,5007, $[$\ion{O}{1}$]~\lambda$6300, and  $[$\ion{S}{2}$]~\lambda\lambda$6718,6732
emission lines gives a spectroscopic redshift of $z_\mathrm{spec}=0.318 \pm 0.009$, with which all
other detected lines are consistent.
This is the first and only known spectroscopic redshift for \xsource,
and we note that it is inconsistent with the photometric redshift of $z_\mathrm{phot}\approx0.17$
as reported in \citet{burgessb} and quoted by the NASA/IPAC Extragalactic Database (NED).
The measured Balmer decrement (H$\alpha$ / H$\beta$ flux ratio)
is $R_\mathrm{obs} \approx 4.42$. Following \citet{tremblay10}, this corresponds to a color excess $E(B-V)$
of
\begin{equation}
E\left(B-V\right)_\mathrm{H\alpha / H\beta} = \frac{2.5 \times \log \left(2.86 / R_\mathrm{obs}\right)}{k\left(\lambda_\alpha\right) - k\left(\lambda_\beta\right)} = 0.437
\end{equation}

where $k\left(\lambda_\alpha\right) \approx 2.63$ and  $k\left(\lambda_\beta\right) \approx 3.71$
as given by \citep{cardelli89}. The excitation
index as defined by \citet{buttiglione2010} is EI$\sim$0.63, consistent with the source being a LEG.
Typical linewidths (e.g., for  $[$\ion{O}{3}$]$ and H$\alpha$) are on the order of $\sim600$ km s\mone.
Fluxes and velocity widths of all $\gae5\sigma$ detected emission lines are given in \autoref{tab:pks1017_spectrum}.

%{\bf Include new optical spectrum for \xsource.
% Pointlike (AGN dominated) or extended in optical?}

\subsection{X-ray imaging and spectroscopy}
\label{sec:results}

Note that because \source\ and \xsource\ are optically classified as LEGs, it is unlikely 
that there is detectable X-ray emission from an accretion disk in these sources \cite[e.g.][]{hardcastle2009}.

\subsubsection{\source}
\label{sec:results1445}

\begin{figure*}
	\centering
	\includegraphics[width=\textwidth]{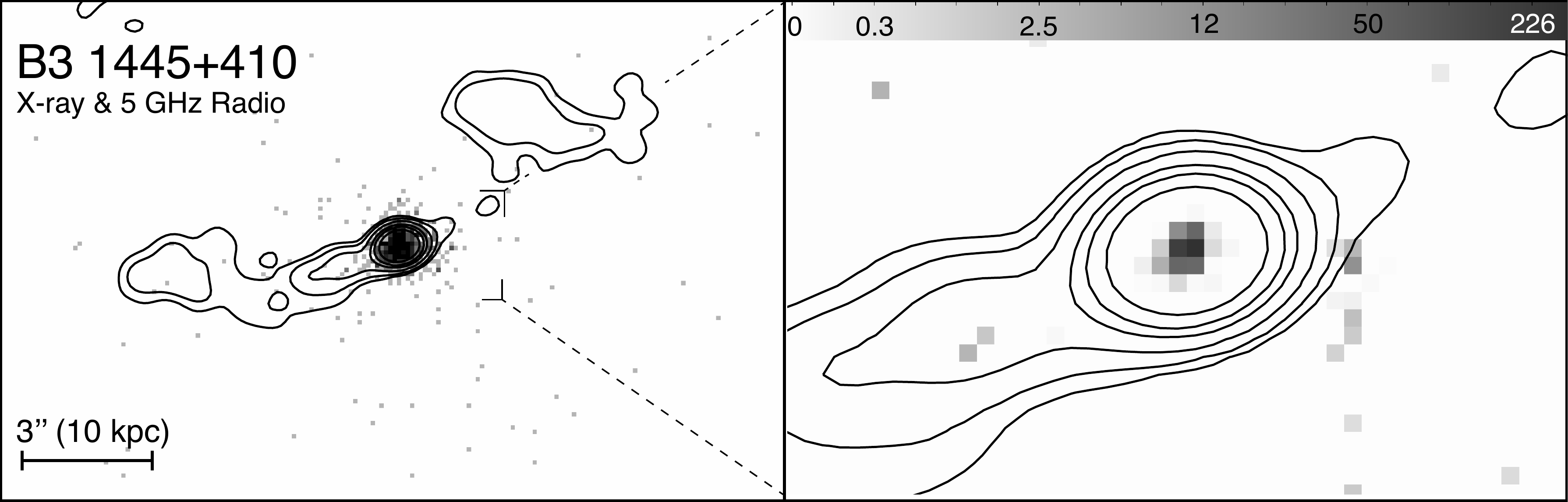}
	\caption{({\it Left}) \chandra\ $ 0.4-5 $  keV data for \source\ with 5-GHz radio
  contours. The \chandra\ data are subsampled into pixels of 0.0984
  arcsec (1/5 of the native detector pixel size).
The radio contours are at 0.3,0.5,1,2,3,5 mJy beam$^{-1}$ and the
restoring beam is $0\farcs57 \times 0\farcs37$. The core flux density is
roughly 26 mJy at 5 GHz. ({\it Right}) Same as for the left panel, with a zoom into the
  central region, where the X-ray data have been deconvolved with the \chandra\
  PSF.
The scale is in units of X-ray counts.}
	\label{fig:1445xradio}
\end{figure*}

\autoref{fig:1445xradio} ({\it Left}) is the \chandra\  image with radio contours and
shows the X-ray emission to be concentrated around the radio core with
no obvious emission from (or structure induced by) the extended radio
features.  To test for spatial extent we modeled the X-ray PSF
appropriate for the observation using the CXC SAOsac raytrace and MARX software.
50 individual simulations were added and the result compared
with the data.  \autoref{fig:1445xradio} ({\it Right}) shows the X-ray data
after using the CIAO {\sc arestore} task to remove the PSF blurring
using the modeled PSF and the Richardson Lucy deconvolution
algorithm. A point source is recovered.

\begin{figure}
	\centering
	%emulateapj%
	\includegraphics[width=0.9\columnwidth,clip=true]{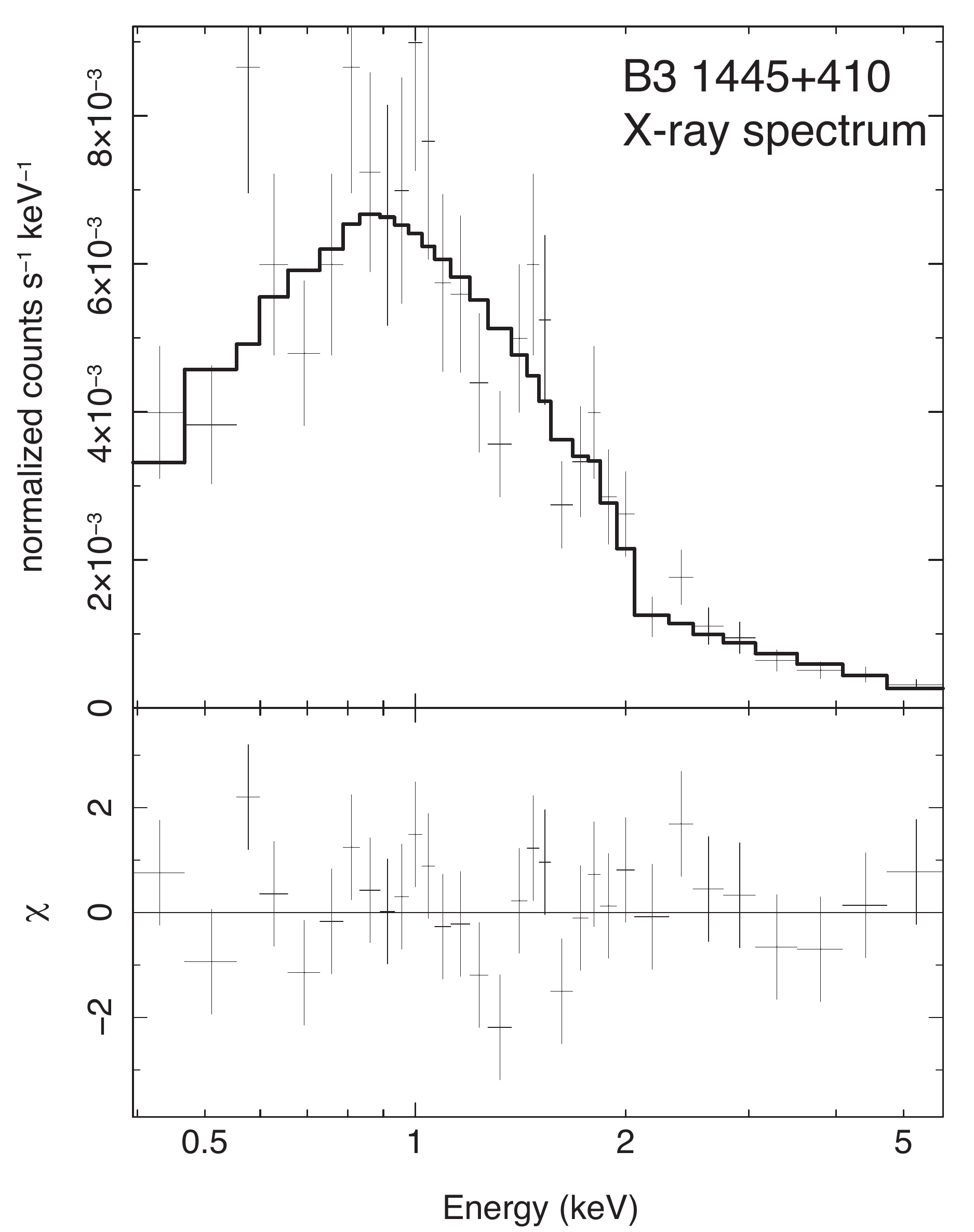}
	%aastex
	%\epsscale{0.67}
	%\plotone{1445corespec.ps}
	\caption{\chandra\ core spectrum of \source\ fitted to a power-law model.  The
		lower panel shows residuals expressed as their individual
		contributions, $\chi$, to $\chi^2$.
	}
	\label{fig:1445corespec}
\end{figure}

The X-ray spectrum (\autoref{fig:1445corespec}) gives a good fit
($\chi^2 =  29.6$ for 30 degrees of freedom) to a power law of X-ray spectral
index $\Gamma = 1.67\pm 0.11$ (90\% confidence), where $\Gamma = \alpha+1$
and flux density is given by $S \propto \nu^{-\alpha}$.   No
  absorption in excess of Galactic is required.  The 1-keV
flux density is $8.9\pm 0.6$ nJy, and between 0.4 and 6 keV the
observed flux is $(6.8 \pm 0.5) \times 10^{-14}$ ergs cm$^{-2}$
s$^{-1}$ and the luminosity is $(7.2 \pm 0.5) \times 10^{42}$ ergs
s$^{-1}$ (90\% confidence uncertainties for one interesting
parameter).

The observation that the radio core is flat spectrum and relatively bright and there is
a one-sided radio jet is consistent with there being some relativistic boosting.
Thus, the measurement that the X-rays are point-like and consistent with a power-law
spectrum suggests that the X-ray emission is non-thermal emission from the radio
core and/or base of the jet.

\subsubsection{3C\,268.3}
\label{sec:results3c268}

\begin{figure*}
	\centering
\includegraphics[width=\textwidth]{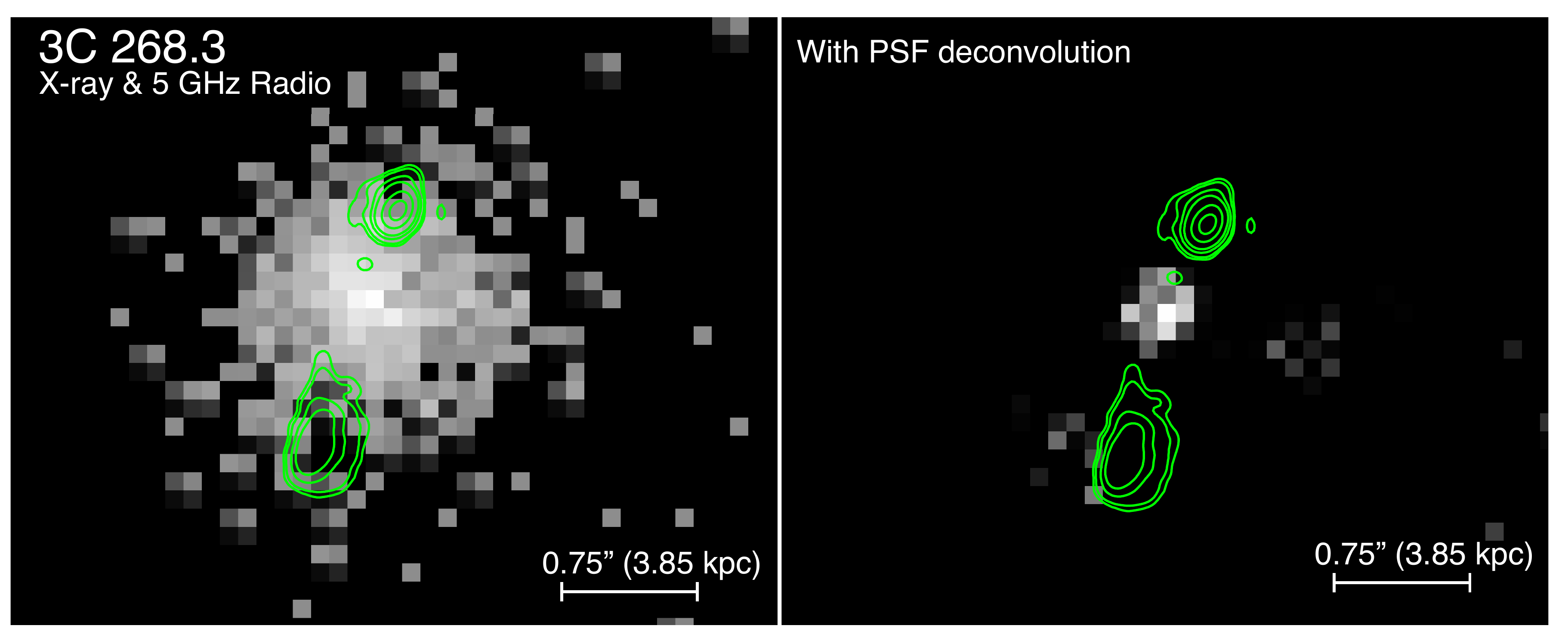}
	\caption{3C\,268.3.  \chandra\ 0.5-5 keV data, with green contours at 0.4, 1, 4, 10, 40, 100
  mJy/beam from the MERLIN 5-GHz map of \citet{ludke1998}  downloaded
  from the NASA Extragalactic Database (NED).
  The \chandra\ data are subsampled into pixels of 0.0984 arcsec (1/5 of
  the native detector pixel size). Left: \chandra\ image.  Right: X-ray
  data after deconvolution with the \chandra\ PSF.  Scales are in X-ray counts.
	}
	\label{fig:3C268chandra_radio}
\end{figure*}

The X-ray emission appears unresolved on inspection. We confirmed this
by simulating a PSF using the CXC SAOsac RAYTRACE and MARX software,
and using the CIAO ARESTORE task to apply the Richardson Lucy
deconvolution algorithm, as used to show that the
CSS source \source\  is also compact in X-rays.  The model PSF was
constructed using the spectral distribution of counts found in the
data (see below).  The X-ray results before and after deconvolution
are shown in \autoref{fig:3C268chandra_radio},  along with contours from a 5-GHz MERLIN radio map
of the source.  A small astrometric correction of 0.17 arcsec was
applied to the X-ray data to centre the source at 12:06:24.715,
64:13:36.98, which is the unconfirmed position of the radio core given
by \citet{ludke1998}.

\begin{figure}
	\centering
	\includegraphics[width=0.5\textwidth]{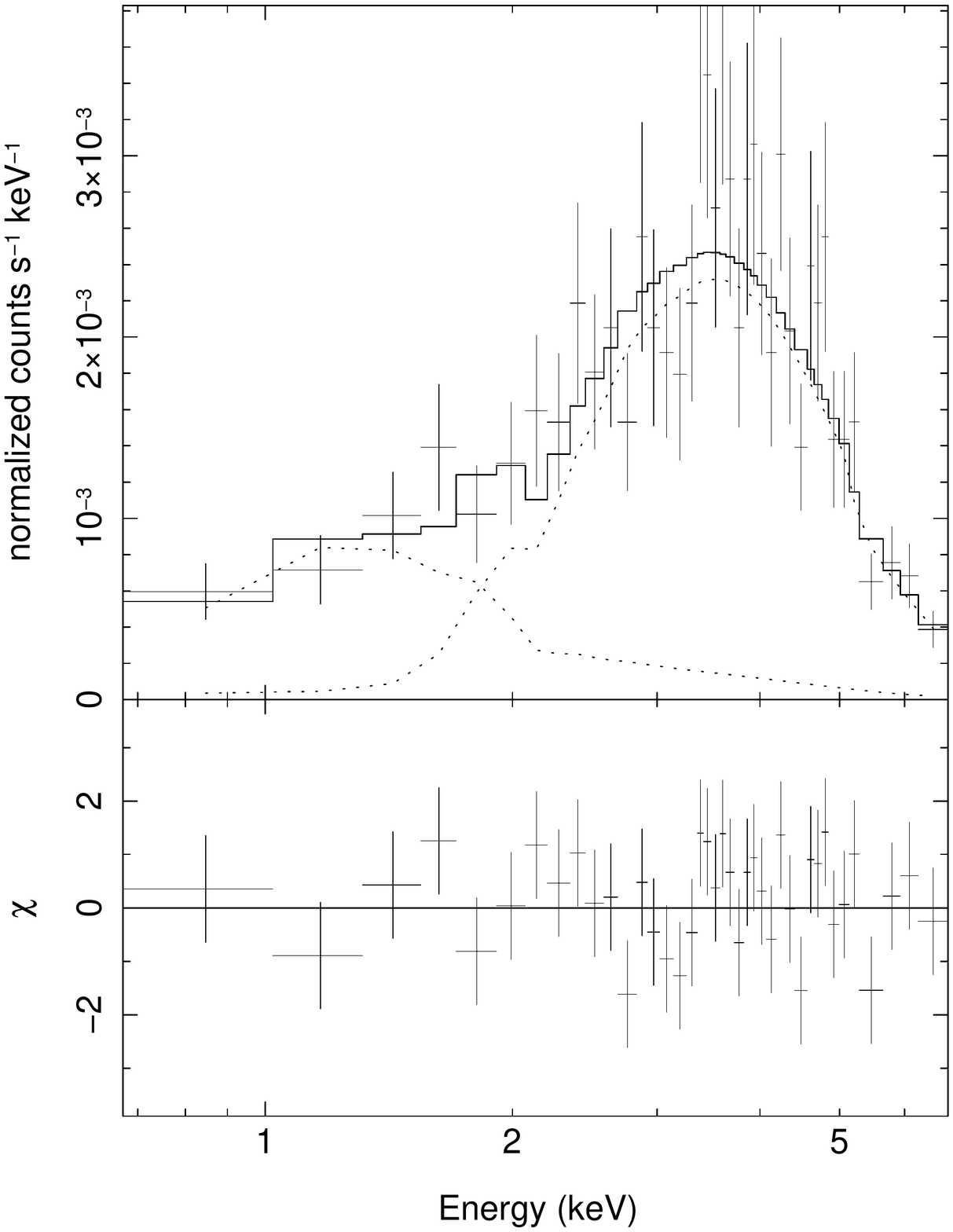}
	\caption{3C\,268.3.  \chandra\ spectrum fit to two absorbed power laws as described
     in the text.}
	\label{fig:3C268chandra_spec}
\end{figure}

The spectrum was measured from counts between energies of 0.4 and 7
keV extracted from a circle of radius 1.25 arcsec, with background
from an annulus of radii 5 and 90 arcsec, masking sources in the
background region found by the CIAO wavdetect software.  Counts were grouped to 15 per
spectral bin, and the weighting scheme of \citet{churazov}  was
adopted to provide an improved estimate of the variance in the limit
of small numbers of counts.  A single-component absorbed power-law
model indicated high absorption but gives an unacceptable fit ($\chi^2
= 71.1$ for 37 degrees of freedom (dof)).  An acceptable fit ($\chi^2
= 31.8$ for 35 dof) is found when a second, less absorbed, power law is
added  (\autoref{fig:3C268chandra_spec}).  The slope of the second power law
 is poorly constrained and so
we have fixed it at $\alpha_2$=0.7. The component values, where $S$
is the unabsorbed 1-keV flux density and $L$ is the unabsorbed
0.4-7-keV luminosity, with 90\%-confidence errors for 1 interesting
parameter, are as follows:  

$\alpha_1 = 0.85^{+0.6}_{-0.5}$,
$N_{\rm H-1} = 1.5^{+0.6}_{-0.4} \times 10^{23}$ cm$^{-2}$,
$S_1 =83^{+153}_{-48} $ nJy, and
$L_1 = 2.9 \times 10^{44}$ ergs  s$^{-1}$.

$\alpha_2 = 0.7 $ (fixed),
$N_{\rm H-2} = 6.6^{+8.3}_{-5.7} \times 10^{21}$ cm$^{-2}$
$S_2 = 2.4^{+2.1}_{-1.2} $ nJy,
$L_2 = 8.8 \times 10^{42}$ ergs  s$^{-1}$.

Merely adding a soft thermal component to the heavily-absorbed power
law is not acceptable (the temperature pegs at the upper limit of 64
keV), but we cannot rule out an (expected) small
contribution to the spectrum from 3C\,268.3's ISM. 

\subsubsection{\xsource}
\label{sec:results1017}

\begin{figure}
	\centering
	\includegraphics[width=0.48\textwidth]{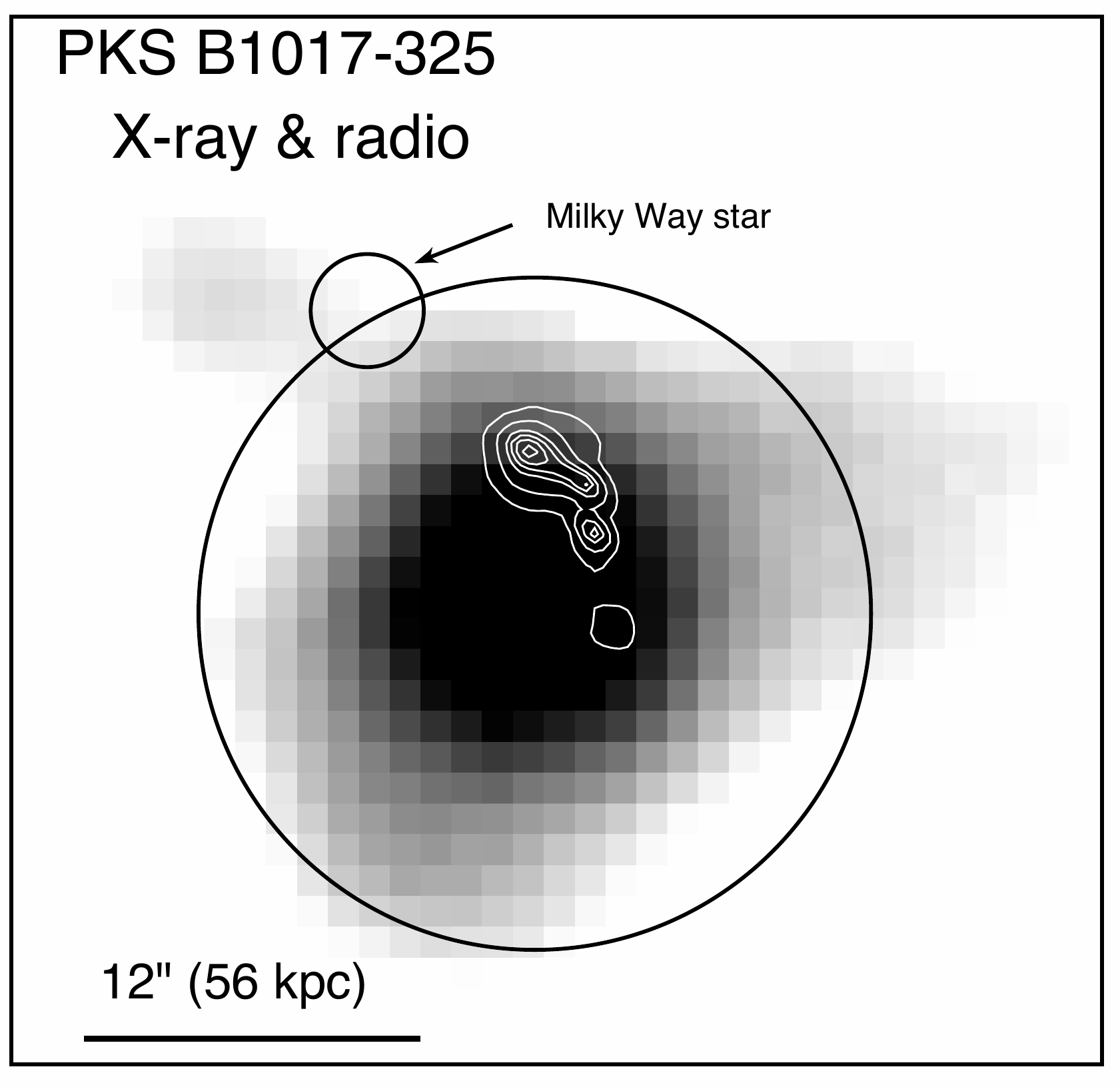}
	\caption{0.4-5-keV \xmm\ image of \xsource\ in 1-arcsec bins and smoothed with
		a Gaussian of $\sigma = 3''$.  The large circle is of radius $12''$
		and the small circle marks the location of USNO star
		U0525\_12970154.
		The white radio contours from our 22.5-GHz map are at
                1, 3, 6, 8, 10, 12
		mJy beam$^{-1}$ and the map has a restoring beam of $1.1 \times 0.6$
		arcsec.
		The core flux density is roughly 9 mJy at 22.5 GHz.
	}
	\label{fig:1017xradio}
\end{figure}

X-ray emission is seen in each of the three \xmm\ EPIC cameras, centered within
1.5~arcsec of the location of \xsource.  \autoref{fig:1017xradio}
shows the X-ray image with radio contours.
Although the X-ray source appears to be centered slightly to the SE of the radio nucleus, this
is within  \xmm's astrometric uncertainties of a few
arcseconds, as confirmed by  \citet{watson2009} during the creation of the
2XMM source catalog.  The X-ray source is  compact,  essentially lying within the circle of radius $12''$ we used for spectral extractions.   The data show no excess centered on the
location of foreground star (U0525\_12970154 in  the USNO-A2.0
catalog) with B = 11.9~mag and r=13.8~mag (marked by the small
circle in  Fig.~\ref{fig:1017xradio}), implying an X-ray to optical flux ratio
less than  about $3 \times 10^{-5}$, consistent with findings for stars
that are this blue in color   \citep[e.g.,][]{vaiana}.

\begin{figure}
	\centering
	%emulateapj%
	\includegraphics[width=0.9\columnwidth,clip=true]{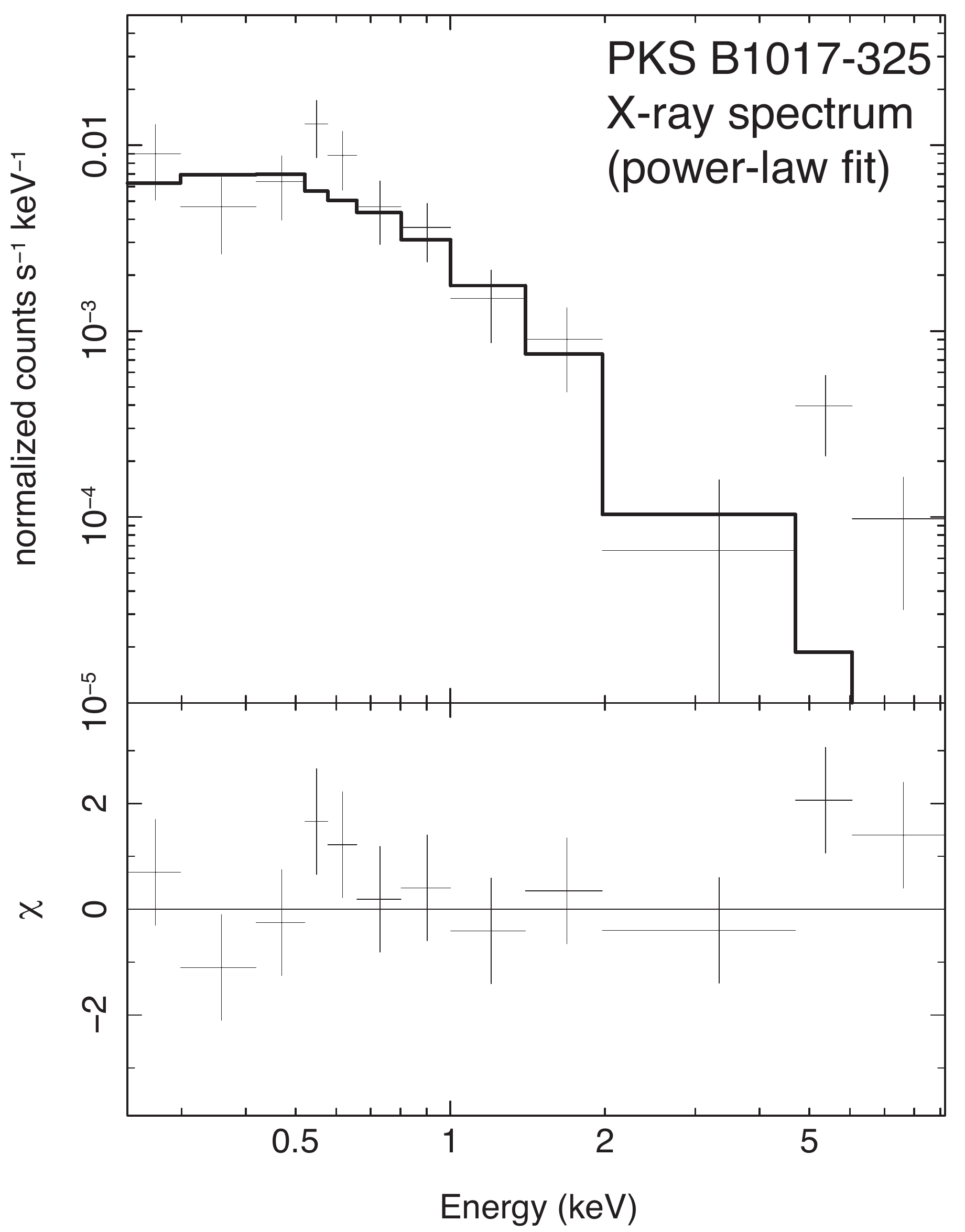}
	%aastex
	%\epsscale{0.67}
	%\plotone{1017corespec.ps}
	\caption{\xmm\ core spectrum of \xsource\ fitted to a power-law model.  The
		lower panel shows residuals expressed as their individual
		contributions, $\chi$, to $\chi^2$.  For clarity of display only the
		pn data are shown.
	}
	\label{fig:1017corespec}
\end{figure}

The X-ray spectrum (\autoref{fig:1017corespec}) gives a good fit ($\chi^2 =
13.0$ for 13 degrees of freedom) to a power law of spectral index
$\Gamma = 2.83\pm 0.36$ (90\% confidence), where $\Gamma = \alpha+1$
and flux density is given by $S \propto \nu^{-\alpha}$.  The 1-keV
flux density is $2.5\pm 0.6$ nJy, and between 0.4 and 6 keV the
observed flux is $(1.1 \pm 0.2) \times 10^{-14}$ ergs cm$^{-2}$
s$^{-1}$ and the luminosity is $(5.8 \pm 1.1) \times 10^{42}$ ergs
s$^{-1}$ (90\% confidence uncertainties for one interesting
parameter).

Since the spectral slope is unusually steep for a radio-galaxy
core \cite[e.g.,][]{evans,massaro2010,massaro2012,massaro2013}, and the
\xmm\ PSF  large enough to contain a significant galaxy atmosphere, we
have also explored fits  to a thermal ({\sc apec}) model.  We
  fixed the abundances to Solar, since  values close to that are
  confirmed observationally for early-type galaxies \citep[e.g.,][]{konami2014} and there
   is no evidence in luminosity or extent for a  significant outer atmosphere where lower abundances are  typical.
A single component thermal gives a poor fit ($\chi^2 = 25.1$ for 13 degrees of freedom),
and adding a thermal to the power-law model finds essentially the power-law component
alone as it pushes $kT$ to its maximum bound of 64 keV.  When we fix
  the  power-law index to $\Gamma = 1.7$, the value for \source, the best-measured X-ray
spectral index of any CSS radio galaxy dominated by core emission, the two-component
model finds an acceptable fit of $\chi^2 = 10.2$ for 12 degrees of freedom.  Here
  $kT =0.42_{-0.13}^{+0.23}$ keV. The 0.4--6-keV luminosities in the thermal and
power-law components are $(2.8 \pm 1.3) \times 10^{42}$ and $(2.9 \pm 1.0) \times
10^{42}$ ergs s$^{-1}$, respectively.  The  1-keV flux density of the power-law
component is $1.2\pm 0.4$ nJy.
 Alternatively, a model with two thermal  components of
solar abundances finds a good fit ($\chi^2 = 9.7$ for 11 degrees of
freedom) with values for $kT$ of $0.39_{-0.13}^{+0.23}$
 and $3.8_{-2.1}^{+u}$ keV, where the upper limit of the hotter thermal is
unconstrained.  The 0.4--6-keV luminosities are comparable in the two
components, at $(2.7 \pm 1.3) \times 10^{42}$ and $(2.7 \pm 1.0)
\times 10^{42}$ ergs s$^{-1}$, respectively.
The cool component, which would be associated with unshocked ISM,
has a temperature consistent with the galaxy
atmospheres of isolated ellipticals hosting nearby low-power radio sources, such as
NGC\,315 \citep{wbh}, although with a luminosity roughly a factor of
ten higher than expected.  Mixed AGN emission could be responsible for
some of the excess luminosity, or
this may be a case where central collapse has both elevated the X-ray
luminosity and triggered the young radio source.
Based on the Rankine-Hugoniot conditions for a strong shock
\citep[see equations for example in][]{wbrev}, the hot component
would be consistent with arising from about 8\% of the ISM mass in
about 2\% of the volume that has been shocked by a Mach number of about 5.
While this would seem a credible model, the sensitivity of the data
limits drawing definitive conclusions.

\begin{figure}
	\centering
	%emulateapj%
	\includegraphics[width=0.9\columnwidth,clip=true]{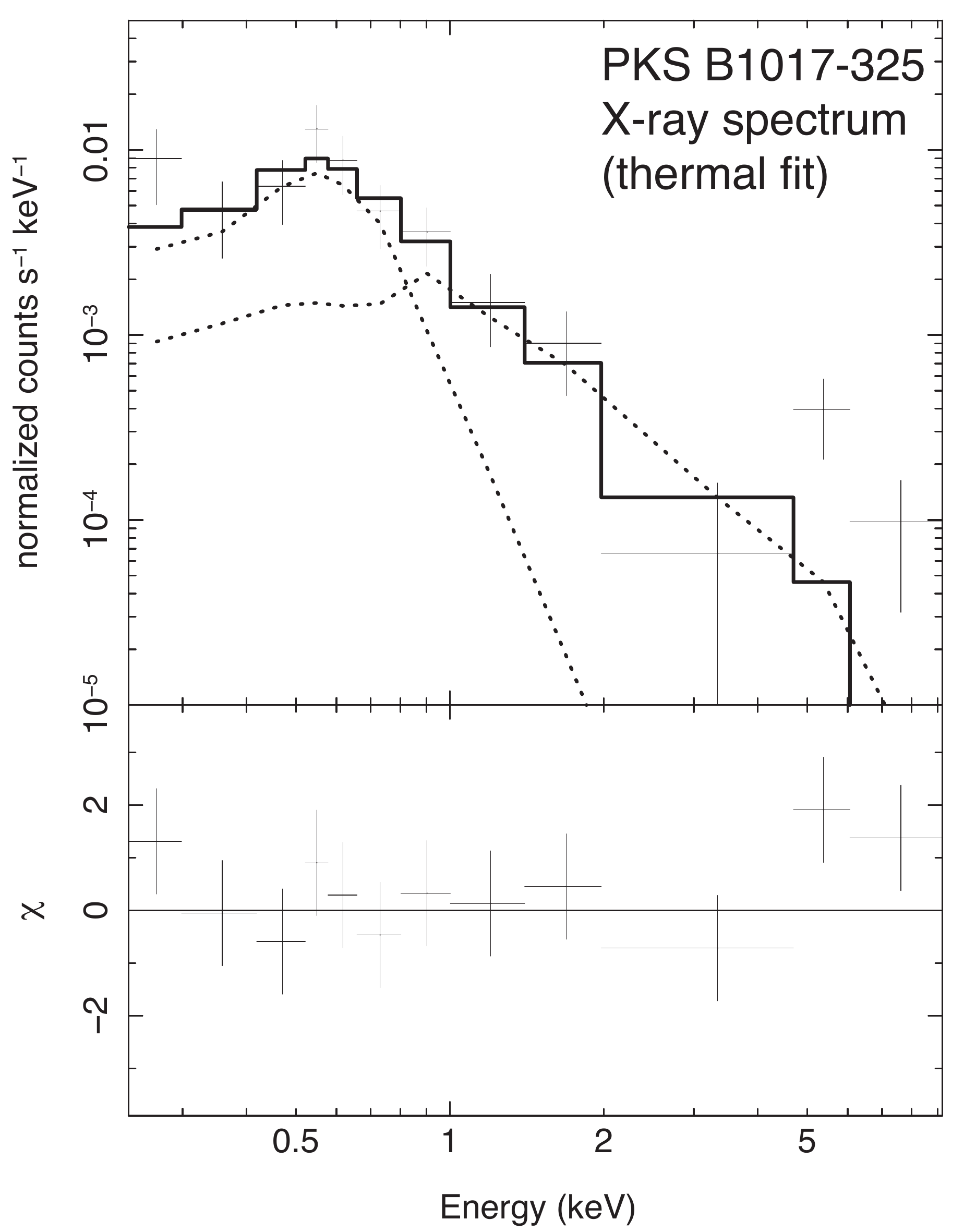}
	%aastex
	%\epsscale{0.67}
	%\plotone{1017apecapec.ps}
	\caption{\xmm\ core spectrum of \xsource\ fitted to two thermal
		models.  Dotted lines show the individual model contributions to the
		fit. The
		lower panel shows residuals expressed as their individual
		contributions, $\chi$, to $\chi^2$.  For clarity of display only the
	pn data are shown.}
	\label{fig:1017apecapec}
\end{figure}

\subsection{Comparison with other CSS Sources}
\label{sec:compare}
\begin{figure*}
	\centering
\includegraphics[width=\textwidth,clip=true]{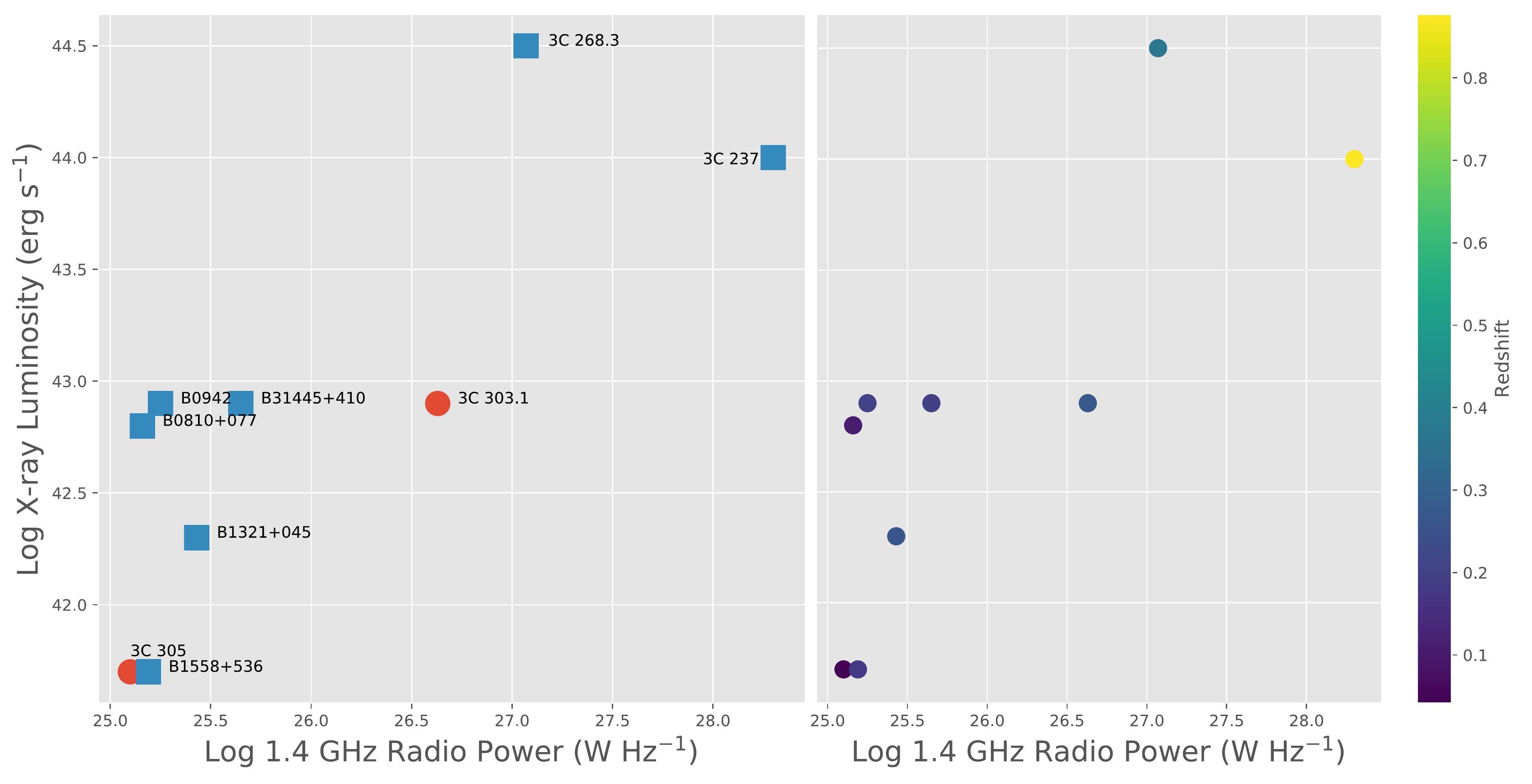}
	\caption{Log X-ray luminosity vs. Log 1.4 GHz radio power for the 9 CSS sources in \autoref{tbl-xray}.
		(Left.) The sources with no X-ray spectral evidence for hot shocked gas are plotted as blue squares, and those with evidence for hot shocked gas are plotted as red circles. (Right.) Same data, but color coded for redshift using scale on right.
	}
	\label{fig:xray-rad}
\end{figure*}

In \autoref{tbl-xray} we present selected properties of the nine known CSS  galaxies with detections in the X-ray\footnote{ Based on our new spectroscopic redshift, the candidate CSS source \xsource\ is now known to have a projected linear diameter of 44 kpc which is significantly above the cutoff for CSS sources. Thus, \xsource\ is not considered a CSS for the  analysis in this section.}.  The counts are generally small. Even so, there is X-ray spectroscopic evidence for hot shocked gas in two of the nine CSS galaxies - 3C303.1 \citep{odea06} and 3C305 \citep{massaro2009,hardcastle2012}.  In addition,  3 of the CSS sources show X-ray emission elongated along the radio source - 3C237 \citep{massaro2017}, 3C303.1 \citep{massaro2010}, and 3C305 \citep{massaro2009,hardcastle2012}. \autoref{fig:xray-rad} shows the X-ray luminosity vs. radio power for the nine CSS sources with X-ray detections (data taken from \autoref{tbl-xray}).  We see an increase in X-ray luminosity with radio power as found by  \citet{Kunert2014}. The Pearson coefficient for this correlation is 0.802, and the associated p-value is 0.009.
A partial correlation analysis between the X-ray and radio luminosity, controlling for redshift as a third variable,  reveals that the two luminosities remain positively correlated. Redshift is nevertheless also a driver of the correlation (\autoref{fig:xray-rad}), and, as with any luminosity vs. luminosity plot (subject to Malmquist and "bigger is bigger" biases), caution should be used in the interpretation of this result.

The CSS radio galaxies with evidence for hot shocked gas are not exceptional in their radio or X-ray luminosities compared to the total sample of nine. We speculate that hot shocked gas is typical in CSS radio galaxies and that deeper X-ray observations will reveal it.

\section{Conclusions}
\label{sec:conclude}

We present X-ray, radio, and optical observations of three  radio galaxies.  3C\,268.3 is a broad line
	CSS radio galaxy.
\source\ is an LEG CSS radio galaxy with possibly a hybrid FRI/II (or Fat Double) radio morphology. The \chandra\ observations are point-like and well fit with a power-law consistent with emission from a Doppler boosted  core.
The \chandra\ observations of 3C\,268.3 are consistent with a point source centered on the nucleus
	whose spectrum can be  fit by two power-law components. 
\xsource\ is a low excitation emission line radio galaxy with a bent double radio morphology.
The \xmm\ observations are
consistent  with ISM emission with either a contribution from hot shocked gas or non-thermal jet emission.
Using  our new spectroscopic redshift, the projected linear size for  PKS B1017-325 falls outside the
       formal definition of a CSS, and thus, we drop \xsource\ from consideration as a CSS source.
 We compile selected radio and X-ray properties of the nine {\it bona fide\/} CSS radio galaxies with X-ray detections so far. We find that 2/9 show X-ray spectroscopic evidence for hot shocked gas and 3 CSS sources show X-ray emission aligned with the radio source.  We note that the counts in the sources are low and the properties of the 2 sources with evidence for hot shocked gas are typical of the other CSS radio galaxies. We suggest that hot shocked gas may be typical of CSS radio galaxies due to their propagation through their host galaxies.

\facilities{CSO, LBT, VLA, XMM}

\acknowledgments

This work was partially supported by NASA grants NNX08AX40G and
GO0-11125X. G.R.T.~acknowledges support from
the National Aeronautics and Space Administration (NASA) through
Einstein Postdoctoral Fellowship Award Number PF-150128, issued by the
\textit{Chandra} X-ray Observatory Center, which is operated by the
Smithsonian Astrophysical Observatory for and on behalf of NASA under
contract NAS8-03060. Basic research in radio astronomy at the Naval
Research Laboratory is supported by 6.1 Base Funding.
The National Radio Astronomy Observatory (NRAO) is operated by Associated Universities Inc. under cooperative agreement with the National Science Foundation.

This work also contains data obtained at the Large
Binocular Telescope (LBT), for which we gratefully acknowledge LBT
Director Dr.~Christian Veillet. The LBT is an international
collaboration among institutions in the United States, Italy and
Germany. LBT Corporation partners are: The University of Arizona on
behalf of the Arizona university system; Istituto Nazionale di
Astrofisica, Italy; LBT Beteiligungsgesellschaft, Germany,
representing the Max-Planck Society, the Astrophysical Institute
Potsdam, and Heidelberg University; The Ohio State University, and The
Research Corporation, on behalf of The University of Notre Dame,
University of Minnesota and University of Virginia.

We also use data
from the Sloan Digital Sky Survey (SDSS).  Funding for the SDSS and
SDSS-II has been provided by the Alfred P. Sloan Foundation, the
Participating Institutions, the National Science Foundation, the
U.S. Department of Energy, the National Aeronautics and Space
Administration, the Japanese Monbukagakusho, the Max Planck Society,
and the Higher Education Funding Council for England. The SDSS Web
Site is \url{http://www.sdss.org/}.

This publication
makes use of data products from the Two Micron All Sky Survey, which
is a joint project of the University of Massachusetts and the Infrared
Processing and Analysis Center/California Institute of Technology,
funded by the National Aeronautics and Space Administration and the
National Science Foundation."

\bibliography{references}

\end{document}